# The lost academic home: institutional affiliation links in Google Scholar Citations[1]

Enrique Orduna-Malea
Universitat Politècnica de València (UPV); Valencia; Spain.

Juan M. Ayllón
Universidad de Granada (UGR); Granada; Spain.

Alberto Martín-Martín
Universidad de Granada (UGR); Granada; Spain.

Emilio Delgado López-Cózar
Universidad de Granada (UGR); Granada; Spain.

## Abstract

**Purpose**
Google Scholar Citations (GSC) provides an institutional affiliation link which groups together authors who belong to the same institution. The purpose of this work is to ascertain whether this feature is able to identify and normalize all the institutions entered by the authors, and whether it is able to assign all researchers to their own institution correctly.

**Design/Methodology/Approach**
Systematic queries to Google Scholar Citations' internal search box were performed under two different forms (institution name and institutional email web domain) in September 2015. The whole Spanish academic system (82 institutions) was used as a test. Additionally, specific searches to companies (Google) and world-class universities were performed to identify and classify potential errors in the functioning of the feature.

**Findings**
Although the affiliation tool works well for most institutions, it is unable to detect all existing institutions in the database, and it is not always able to create a unique standardized entry for each institution. Additionally, it also fails to group all the authors who belong to the same institution. A wide variety of errors have been identified and classified.

**Research Limitations/Implications**
Even though the analyzed sample is good enough to empirically answer the research questions initially proposed, a more comprehensive study should be performed to calibrate the real volume of the errors.

**Practical implications**
The discovered affiliation link errors prevent institutions from being able to access the profiles of all their respective authors using the institutions lists offered by Google Scholar Citations. Additionally, it introduces a shortcoming in the navigation features of Google Scholar which may impair web user experience.

**Social implications**
Some institutions (mainly universities) are under-represented in the affiliation feature provided by Google Scholar Citations. This fact might jeopardize the visibility of institutions as well as the use of this feature in bibliometric or webometric analyses.

**Originality/Value**
This work proves inconsistencies in the affiliation feature provided by Google Scholar Citations. A whole national university system is systematically analyzed and several queries have been used to reveal errors

---

in its functioning. The completeness of the errors identified and the empirical data examined are the most exhaustive to date regarding this topic. Lastly, some recommendations about how to correctly fill in the affiliation data (both for authors and institutions) and how to improve this feature are provided as well.

# 1. Introduction

Google Scholar Citations (GSC) is the academic profile service created by Google (now Alphabet) in 2011, based on the bibliographic data available in Google Scholar (Jacsó, 2012). Among its main features is the complete freedom users have to edit their personal information (name, institutional affiliation, and areas of interest). Faithful to the philosophy of the company, a loss of precision in affiliations and subject categorizations is expected because of the absence of any terminology constraints, which, on the upside, allows a more flexible and open experience. This is in contrast to other, more constrained profile services, like the one Microsoft Academic Search used to offer (Jacsó, 2012).

However, in August 2015 Google Scholar Citations introduced a new feature aimed to facilitate affiliation searches and browsing the profiles of authors who work in the same institution. Thus, if an author has entered an affiliation in a more or less conventional form, a link will automatically appear pointing to the list of all researchers working in the same institution, while also displaying the usual information available in a Google Scholar Citations search: total number of citations received by each researcher, areas of interest entered by the author (up to five of them). This feature is therefore an evident improvement in the product, since it offers a simple a quick method to search authors by their institution. Although these searches could already be carried out before (by searching the domain of the institutional email), the process was too slow and tedious to be used for large-scale quantitative studies, and excluded authors using unofficial e-mail addresses.

This new functionality not only facilitates the identification of all authors working in a university or in other organizations (providing their profiles are public), but is also useful if you want to learn about the scientific interests and subject profile of an institution. This fact may, indirectly, make it easier to perform institution evaluation processes (especially university rankings), as well as to evaluate authors from the same institution, as it has already been done – in one way or another – in other profile platforms like the already mentioned Microsoft Academic Search, and more recently, ResearchGate (Thelwall & Kousha, 2015). The Top universities by Google Scholar Citations (http://webometrics.info/en/node/169) and the Ranking of researchers by country (http://webometrics.info/en/node/116), both developed by the Cybermetrics Lab at the Spanish National Research Council (CSIC), serve as examples of this kind of initiatives.

It is for this reason that the high or low accuracy with which institutions are normalized can have a major impact in the visibility of institutions in this particular product, and therefore, in the potential institution-level studies – with evaluative purposes or otherwise – that may be carried out using these data, as it happens with other bibliographic products (García-Zorita et al, 2006; Venets, 2014).

The identification and unification of institutions is a difficult and intricate task, full of complex and unexpected cases, mainly due to the wide variety of forms researchers use to enter their institutional affiliations (uncommon variations, spelling errors or even complete absence of information). It's even worse in non-English speaking countries, where institutions can be entered in their original form, or translated to English. These issues affect international visibility of universities negatively, making bibliometric studies based on affiliations and university rankings troublesome (Bador and Lafouge, 2005; Van Raan, 2005; Praal et al, 2013; Taşkın and Al, 2014).

Huang et al (2014) differentiate between two kinds of problems in institutional affiliations: MTS (Multiple to simple) when various institutions share the same name; and STM (Simple to Multiple), when one institution can be referred to by several names. Some of the reasons why the latter phenomenon may occur are translation, spelling, institutions that change their name, errors, and divisions. These problems are exacerbated when there is no authority control in the data entry process, and the system gives users the freedom to identify themselves as they like. The main methods to fix these errors and reach a complete unification of the variants are clustering through word similarity, and using variant lists developed by competent authorities. To date, the literature has proposed quite a diverse automatic institutional name disambiguation and normalization processes (Galvez and Moya-Anegón, 2006; 2007; Jiang et al, 2011; Cuxac, Lamirel and Bonvallot, 2013; Hu et al, 2013; Morillo et al, 2013; Morillo, Santabárbara and Aparicio, 2013; Huang, 2014), thought any of them has completely solved the problem.

These normalization problems can entail a numbness for authors (being not appropriately linked with their institutions), institutions (diminishing their web visibility by not showing all existent authors with a public profile created and affiliated with the organization), researchers and information professionals (managing incomplete authors lists), and science policymakers (using an incomplete information list to deliver decisions or judgements).

For this reason, the purpose of this study is to describe the workings of this new institutional search and browsing feature available in Google Scholar Citations, as well as its effectiveness to identify in a precise way all the authors working at a given institution. Hence, the goal is to detect and categorize its functional limitations, and to propose best practices in assigning author affiliation in the platform. This may help both authors to prevent these errors when editing their profiles, and information researchers and professionals to consider them appropriately when using these authors' lists.

## 2. Related research

Currently, an increasing number of platforms allow researchers to build an academic profile on the Web, including not only their curriculum and professional activities, but also a list of their publications with a variety of impact metric attached to them (based on citations, visits, or mentions in social networks). These profiles can be classified according to the source of data they use.

Thus, academic social networks are one group. Among them we can find ResearchGate (Hoffmann, Lutz, & Meckel, 2015; Thelwall & Kousha, 2015; Orduna-Malea, Martin-Martin & Delgado Lopez-Cozar, 2016; Thelwall, Kousha, in press), Academia.edu

(Thelwall & Kousha, 2014), and Mendeley (Li, Thelwall & Giustini, 2014). These platforms enable authors to create a profile, but at the same time they also facilitate social interaction among researchers. The second group would be profiles created in repositories, whether they are institutional repositories (http://futur.upc.edu), subject-oriented repositories (https://www.ssrn.com) or meta-repositories (http://citec.repec.org). In both cases, the information in these profiles depends directly on the contents that researchers contribute to the social network or repository.

Lastly, there are also academic profiles based on the information available in an academic search engine (Ortega, 2014). These profiles can be created either automatically, like in AMiner (https://aminer.org) or the new Microsoft Academic (https://academic.microsoft.com), or manually, like in Google Scholar Citations (https://scholar.google.com/citations). The main advantage of these profiles is that they have a greater coverage than repositories (institutional or subject-oriented), as well as their better capabilities for retrieving academic content online.

Among the many advantages of academic profiles we can highlight the chance for structured data extraction to be used in bibliometric studies. Authors fill in their academic profiles not only with their publications but also with personal information such as their institutional affiliation. So, specific data from author profiles may be directly used to ascertain institutional level performance, whether it is the presence of institutions in the platform (social population studies) or research impact (bibliometric studies). While some platforms like ResearchGate or Microsoft Academic provide institution-level profiles natively, other platforms require manual and/or automated data extraction to be able to generate statistics about affiliations. That was precisely the case of Google Scholar before the launch of Google Scholar Citations.

As a consequence of this, affiliation studies heavily depended on the accuracy of Google Scholar's raw data. Given the lack of data exporting functionalities, the absence of authority control, and the little interest in recovering publishers' bibliographic metadata, the amount of bibliographic errors were remarkable (Jacsó, 2009; 2010; 2012a), errors of which traditional bibliographic databases were not exempt either, although to a lesser extent (Franceschini, Maisano & Mastrogiacomo, 2016a; 2016b; Tüür-Fröhlich, 2016). Some authors even requested alternative methods to conduct analysis of the presence of institutions in Google Scholar via webometric methods (Aguillo, 2012).

The launch of Google Scholar Citations (Butler, 2011), which provided structured affiliation data and allowed users to edit their personal information somewhat improved the situation, opening a new scenario for analyzing bibliographic data from Google Scholar (Huang & Yuan, 2012).

Since then, Google Scholar Citations has been widely analyzed in the scientific literature. There are studies that examine its basic functionalities, size, contents (authors, countries, universities, and areas of interest), and demographic aspects (Jacsó, 2012b; Ortega & Aguillo, 2012; 2013; 2014; Ortega, 2015a; 2015b). Regarding its use at the institutional level, it is worth to mention the analysis of the Spanish National Research Council (Ortega, 2015c) and the University of Bergen (Mikki et al, 2015), where data from GSC were retrieved and compared with several altmetric sources. At the discipline level, Martín-Martín et al. (2016) analyzed the area of Bibliometrics and Scientometrics using Google Scholar Citations in combination with other platforms (ResearchGate, ResearcherID, Mendeley, and Twitter) whereas Ortega (2015d) attempted to define which research attributes (gender, academic position and discipline) bring together more successful profiles through the analysis of more than 3000 profiles from Google Scholar Metrics.

However, despite the clear advantages that Google Scholar Citations provides for carrying out bibliometric analysis, the lack of authority control still diminishes its usefulness. For example, Ortega (2015b) claims that, for instance, more than 20 variants to refer to the Universidade de São Paulo were found, which requires a "hard manual data cleaning". The launch of the Google Scholar Citations' affiliation link is therefore a step to solve this issue.

Recently, Utrecht University researchers Bianca Kramer and Jeroen Bosman released a comprehensive report about the use of academic communication tools (https://101innovations.wordpress.com) based on more than 20,000 responses by researchers and editors to a survey. The results indicate that Google Scholar Citations is used by 62% of the users (in second place after ResearchGate, with 66%), although in combination with Google Scholar, it is the most used tool to set up alerts, recommendations, and to search scientific literature.

Given the intensive use of Google Scholar Citations among researchers worldwide, the effect of the normalized institution feature on the visibility of academic institutions is expected to be high. However, the effectiveness of this feature hasn't been approached before.

*Google Scholar's affiliation feature*

This new feature can be found in the area of the profile where the institutional affiliation of the author is displayed. It is active when the institution entered by the author is displayed as a link. When clicking on this link, the user is taken to a list of all the authors Google Scholar has identified as working for that institution, sorted according to their total number of citations received (Figure 1). Users can browse the authors in the list. The technical requisite for this link to appear is that the institution entered must match the institutional email. If this condition is not met, the link won't be active even if the author entered the name of the institution correctly, and therefore he or she won't appear in the list of authors working for that institution.

**Figure 1. Linked standardized institutional affiliations in Google Scholar Citations. Example for Joseph E. Stiglitz at Columbia University.**

The feature only works for the search box available in Google Scholar Citations. When the search string contains the name of an institution (i.e. California), the system now also suggests a list of institutions containing that term (i.e. University of California, Berkeley), in addition to the list of author profiles in which that term appears. That way, users can access the list of authors working in any given institution (Figure 2). Although it is not possible to use this feature directly from the Google Scholar search interface (not even from the advanced search menu), we acknowledge the appearance of recommended links among the search results when querying for institution names, allowing the user to access directly to the authors list provided by Google Scholar Citations to one specific institution.

**Figure 2. Institutional search by keywords. Example for "California".**

One of the main strengths of the feature is its ability to recognize different variants of the same institution (STM problem), especially the ones caused by translations to various languages. In Figure 3 we can see how after searching in Catalan (Universitat de Barcelona), the system returns the normalized name of the institution in English (University of Barcelona). It also returns author profiles in which the institution appears in Spanish (Universidad de Barcelona). The specific criteria chosen to select which variant should be the normalized one haven't been disclosed, and they are not very stable. As an example, we noticed that the normalized version for the university in Figure 3 was, by September 2015, the Catalan version, although this was later modified to the English version.

**Figure 3. Authority control for institutional name variants in different languages. Example for Universitat de Barcelona.**

In addition, the system also tries to solve the MTS problem, that is, differentiating institutions that consistently use the same acronym. In figure 4 we can observe a search of the acronym "UPV", shared by two Spanish universities, Universitat Politècnica de València (Polytechnic University of Valencia), and Universidad del País Vasco (University of the Basque Country).

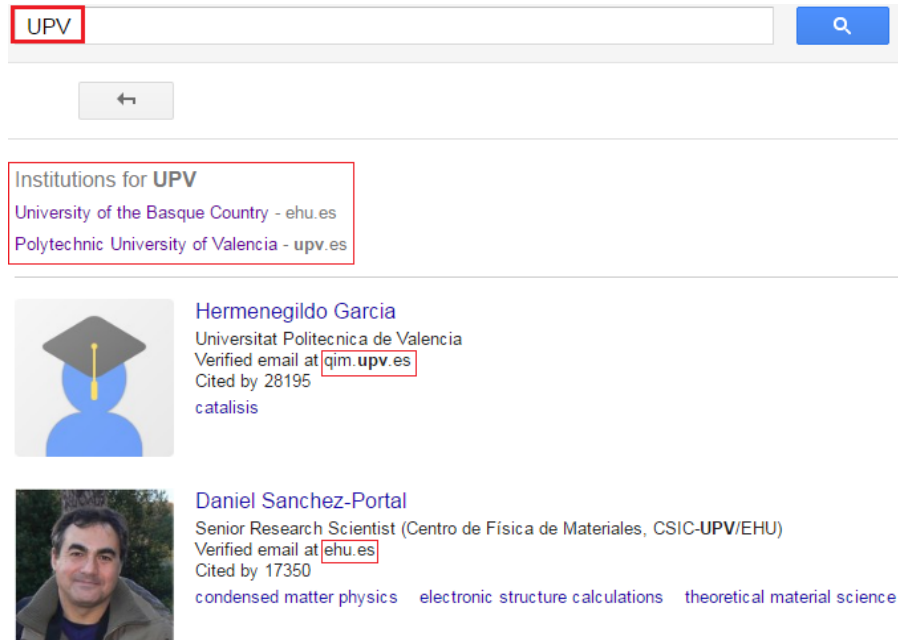

**Figure 4. Authority control for identical acronyms corresponding to different universities. Example for Polytechnic University of Valencia and University of Basque Country.**

## 3. Research questions

The main objective of this study is to analyze the new affiliation feature available in Google Scholar Citations, trying to determine its degree of precision in the task of identifying and classifying authors in their respective institutions. In order to do this we ask the following research questions:

- (RQ1) Is Google Scholar Citations able to identify all institutions mentioned in author profiles?
- (RQ2) Is Google Scholar Citations able to create a unique preferred name representing all variations, including different spellings and misspellings, uppercase versus lowercase variants, acronyms and abbreviations?
- (RQ3) Is Google Scholar able to correctly assign all scholars to their respective academic institutions?

## 4. Methods

Aiming to answer those research questions, we designed a series of queries for the Google Scholar Citations search box with which we tried to uncover errors in the functioning of the affiliation feature. A series of text queries (i.e. "Universidad Complutense de Madrid") were sent to the platform in order to get data for all the institutions that are part of the current Spanish university system. We searched the common different variants of the same university: complete name in Spanish and English, and other regional languages when necessary (i.e. Universidad de Barcelona, Universitat de Barcelona, and University of Barcelona), acronyms (UB), as well as web domains and their variants (ub.cat, ub.edu, ub.es).

The selection of this sample is motivated by the fact that it is an easily manageable system (82 public and private universities), with a marked language diversity (which facilitates testing institutions with several official names), and where there is a central research organization (CSIC) that introduces the possibility of double affiliations (University/CSIC), useful to test the precision of the system when it has to deal with complex affiliations. The use of a whole university system like this as a sample guarantees that the vast majority of inherent errors in automatic affiliation assignment may emerge. The use of other academic systems with their own particularities (such as size or diversity) may reflect different error rates but not different error types.

Additionally, we also carried out searches containing common terms and names of world-class universities in order to observe the general working of the system. As a special case, we sent queries to retrieve profiles from authors working in non-academic institutions (Google in particular) with the goal to complement the results obtained to the third research question (RQ3) with data from organizations other than universities.

The results obtained for each query were analyzed quantitatively to try to identify and categorize the various types of errors and limitations found. All queries were carried out during September 2015.

## 5. Results

*RQ1: Coverage: are all institutions standardized?*

A significant percentage of Spanish universities (16%; 13 in total) don't have a normalized entry in Google Scholar Citations. This happens in spite that for some of these institutions (like Universidad Antonio de Nebrija), authors have correctly entered their affiliation and verified their institutional email. Therefore, the link pointing to the list of authors in the institution is not available for these universities. In other cases (Universidad Alfonso X El Sabio), none of the authors have verified their institutional email, so we can't know for sure whether those universities are correctly recognized by the system. However, the fact that the system doesn't show any normalized version (see figures 2-4) when we carry out a query to find them, makes us believe that not only it is not normalized, but that the system doesn't use automated procedures to identify institutions.

This reveals the absence of an indeterminate number of universities in the system. As an example, we couldn't find the following institutions: Universidad Nacional de Loja y Universidad Católica de Santiago de Guayaquil (Ecuador), Universidad del Valle de México (Mexico), Pontificia Universidad Católica Argentina (Argentina), and Universidade Católica de Pelotas (Brasil). These institutions are obviously not world-class institutions (which are all well covered by GSC), but their absence from the system reveals a problem of coverage in the identification of institutions where authors with a public profile in Google Scholar Citations work, thus limiting the suitability of the platform for bibliometric and cybermetric analyses.

*RQ2: Methodological errors: does the identification algorithm operate properly?*

The mechanism of the feature (match between the name of the institution and the institutional email) works for the most part, although we have observed a lack of flexibility when it comes to detecting common errors made by users at the time they enter their affiliation. For example, the term "unversity" appears in more than 300 profiles around the world. This same error (which also appears in other languages: unversidad, unversität) and similar ones are not taken into account by the system, and therefore these profiles are not included in their respective institutions.

Apart from this limitation, which to be fair is caused by a mistake made by the researchers (although it again indicates the use of a closed list of universities instead of an automated system to identify institutions), we found the following issues:

*a) Wrong normalized name*

In some cases Google Scholar has not chosen as the normalized name the official name of the institution, but some other. For example, the Universidad Católica San Antonio de Murcia (Catholic University San Antonio of Murcia) is identified simply as Universidad Católica de Murcia (Catholic University of Murcia). Moreover, the link is sometimes correctly added to the acronym: UCAM (See Figure 5; right, example A), but not to the correct full name (Figure 5; right, example B). In this case Google Scholar is probably being influenced by the information available in Wikipedia, which is used to construct Google's knowledge graph (Figure 5 left).

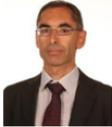

**Figure 5. Selection of incorrect normalized names of institutions.**

*b) Disambiguation problems*

The affiliation link feature does not use topographic qualifiers for discriminating different universities with the same name. This is the case of the Universidad del Valle, same name used by four different institutions in Colombia, Guatemala, Mexico, and Bolivia. In this case, the tool has selected the Colombian institution as the only valid form (https://scholar.google.com/citations?hl=en&view_op=search_authors&mauthors=Universidad+del+Valle).

*c) Incorrect linking of different universities*

In some cases the identification of affiliations is not done correctly. For example, Universidad CEU San Pablo, Universidad CEU Cardenal Herrera, and Universitat Abat Oliva CEU are formally independent universities, although they all belong to the same entity: Centro de Estudios Universitarios – Center for University Studies (CEU).

However, when Universidad Cardenal Herrera is identified as an affiliation (Figure 6; top), it redirects automatically to Universidad CEU San Pablo (Figure 6, bottom). This error may be attributed to the great difficulty of knowing about these individual cases, which depend on the national legislation applicable in each country.

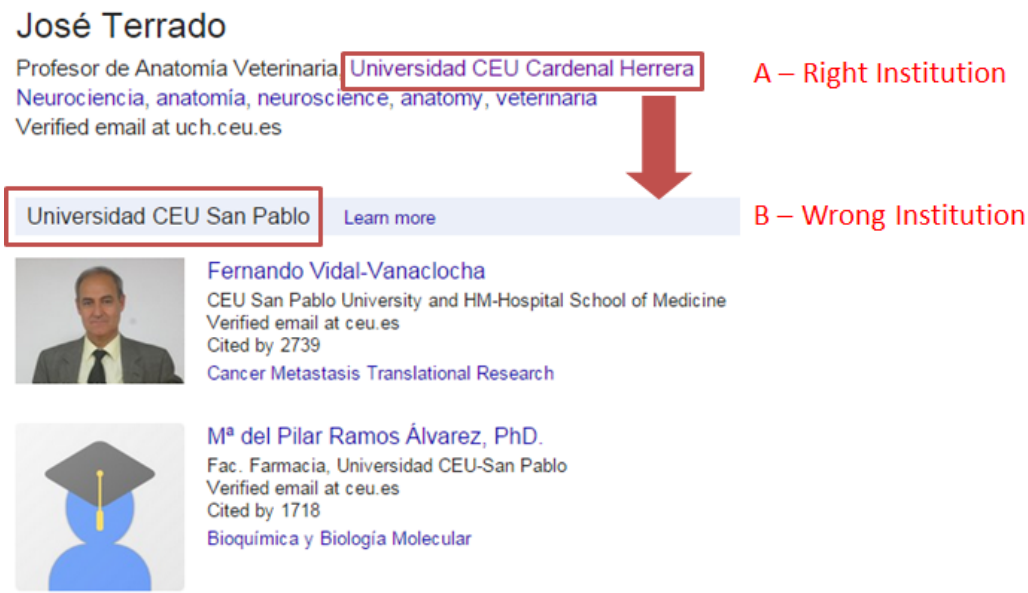

**Figure 6. Combination of different institutions.**

*d) Multiple official academic web domains*

Some universities use different academic web domains for SEO (search engine optimization) purposes, which in the case of universities is considered a bad practice from a webometric point of view (Orduna-Malea, 2012). In these cases, the feature does not work correctly, as only one web domain is considered as valid (that is, web domain variants are not considered institution name variants). Continuing with the previous example, Universidad CEU Cardenal Herrera has two different web domains (uch.ceu.es and uchceu.es). In this case it seems that "uch.ceu.es" is being considered as the valid web domain so all author profiles containing an "uchceu.es" verified email are not correctly linked to the institution. This phenomenon, however, probably won't affect academic systems with standardized university web domains (such as US or UK universities, with .edu or .ac.uk domains, respectively).

*e) Lack of precision of the search box*

Additional problems have also been detected in the search box, which does not work correctly in its task of identifying all the institutions that contain a particular keyword. For example, if the term "Polytechnic" is queried, no institutions are returned (the same happens with the Spanish term "politécnica" or catalan term "Politècnica". Similarly, the term "University" doesn't return any normalized institutions either, whereas the term "Catholic" retrieves three: Catholic University of Korea, Catholic University of America, and Università Cattolica del Sacro Cuore. The last one doesn't even contain the term that was queried, but the Italian translation.

If the term queried is a city or a place (See figure 2 for California, and Figure 7 for Barcelona), the feature does not return all normalized institutions containing the corresponding term. The same thing happens with the major cities in Spain: "Barcelona" returns three normalized institutions (University of Barcelona, Pompeu Fabra University, and Polytechnic University of Catalonia), and Madrid only two (Complutense University of Madrid, Universidad Politécnica de Madrid), when in fact these cities host many more universities, which can in fact be found when they are searched directly by their names (i.e. Autonomous University of Barcelona, or Universidad Autónoma de Madrid). In some cases the normalized institution is the original Spanish one, whereas in others, the English version has been selected.

It seems that the search box only suggests the institutions that may match the query, while also returning author profiles in which the queried term appears in any part of the personal information area. This means that institution searches are only effective when searching specific institutions. Moreover, we didn't get more than three normalized institutions in any of our queries.

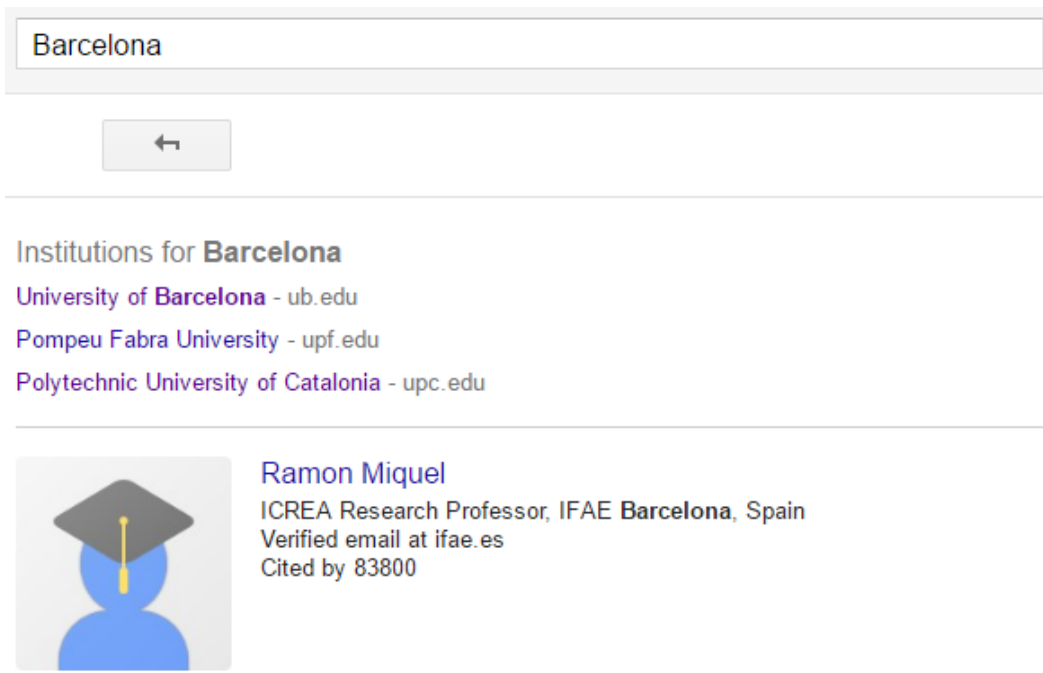

**Figure 7. Finding institutions in the Google Scholar Citations search box.**

*f) Complex institutions*

We encountered various problems with complex institutions. This is the case of the Spanish National Research Council (CSIC) in Spain. Although the standard term in Google Scholar Citations for this institution is "CSIC", mixed centers (ventures between universities and research institutions belonging to CSIC) are not linked correctly.

Some authors enter their affiliation mixing in a variety of forms the acronyms of CSIC and the corresponding institution (for example CSIC-UPV or UAM/CSIC). In these cases, the activation link will be active only for the corresponding institution used in the corporate email. In some cases the university will be the institution that is activated (Figure 8, case A) while in some others, it'll be CSIC (Figure 8, case B). If the acronym has been entered following other more uncommon variants, the link might not be activated at all, even if there is a matching verified email (Figure 8, case C). Lastly, if authors enter their affiliation membership by adding the specific research institute (in the form Research institution – CSIC or vice versa), the link might not be activated either (Figure 8, case D).

**Ismael Rafols** — Case A: University linked
Ingenio (CSIC-UPV), Universitat Politècnica de València; SPRU, University of Sussex
Science and Technology Policy, Evaluation, Scientometrics, Science Technology and Studies
Verified email at ingenio.upv.es - Homepage
https://scholar.google.com/citations?user=puqLKfsAAAAJ

**Alfredo Villasante** — Case B: CSIC linked
Centro de Biología Molecular "Severo Ochoa" (CSIC-UAM)
centromere, telomere, heterochromatin, noncanonical DNA structures
Verified email at cbm.csic.es - Homepage
https://scholar.google.com/citations?user=GzDMYYAAAAAJ

**Juan Jose Hernandez Rey** — Case C: No institution linked (I)
IFIC - Instituto de Física Corpuscular (CSIC and UV), Spain
Particle and Astroparticle Physics
Verified email at ific.uv.es - Homepage
https://scholar.google.com/citations?user=DjXSkTEAAAAJ

**Josep Penuelas** — Case D: No institution linked (II)
Research Professor at Global Ecology Unit CREAF-CSIC Barcelona
global ecology, atmosphere-biosphere interactions, remote sensing, climate change impacts, ecometabolomics and biogenic VOCs
Verified email at uab.cat - Homepage
https://scholar.google.com/citations?user=7uwY4MgAAAAJ

**Figure 8. Affiliation for complex institutions in Google Scholar Citations. The case for mixed University – CSIC research institutions in Spain.**

*g) Multiple affiliations*

Authors with more than one affiliation are another collective of profiles for which the affiliation feature doesn't work correctly. This is the case of Matthew Fujita (https://scholar.google.com/citations?user=JtiRiK8AAAAJ), who entered affiliations to University of Texas at Arlington, University of California Berkeley, Harvard and Oxford. Only one institution can be activated with the affiliation link (in this case Texas at Arlington, because it is the one that matches the email). While this procedure may prevent that sporadic visitors may request false or gamed affiliations in Google Scholar

Citations, all professors and researchers with more than one real affiliation must therefore select only one institution to be included in the normalized lists.

h) *Internal affiliations*

We also noticed that authors who enter sub-affiliations (departments, research institutes, Faculties, Schools, etc.) are not included (Figure 9). This is an issue for prestigious centers that are well-known in their own right. This is the case, for example, for Schools belonging to Harvard, such as the Harvard Medical School (2124 profiles as of May 2016), Harvard Business School (102 profiles) or Harvard Law School (25 profiles). Other schools, however, are correctly identified, such as Harvard School of Public Health (linked with "channing.harvard.edu" email). Probably, the existence of multiple emails from some Schools (childrens.harvard.edu, hms.harvard.edu, joslin.harvard.edu, crystal.harvard.edu, among many others, for Harvard Medical School), due to the existence of different centers belonging in turn to them, makes the feature work incorrectly too in these profiles.

**Figure 9. Authors with internal institutional affiliations.**

**RQ3:** *Precision: are all authors catalogued in their corresponding institution?*

Google Scholar fails to completely group together all the researchers who work in the same institution. The reason for this lies in the criterion used for the "author-organization" link: only those authors who have simultaneously indicated the name of the institution in the affiliation field and have verified their profile with a matching email are added to the list of researchers of the institution.

To illustrate this, we offer two cases studies. The first deals with a private company (Google), and the second one with the set of Spanish universities.

*a) The case of a company: Google*

To illustrate this, consider the profiles of the two fathers of Google Scholar (Anurag Acharya and Alex Verstak). As we can see in Figure 10, Anurag Acharya's profile is correctly included while Alex Verstak's profile is not linked to any institution. In the case of Acharya's profile, both the institutional affiliation field and the verified email are a match for the same company: Google. In contrast, the affiliation declared by Verstak is Virginia Tech whereas the verified email is from Google (google.com).

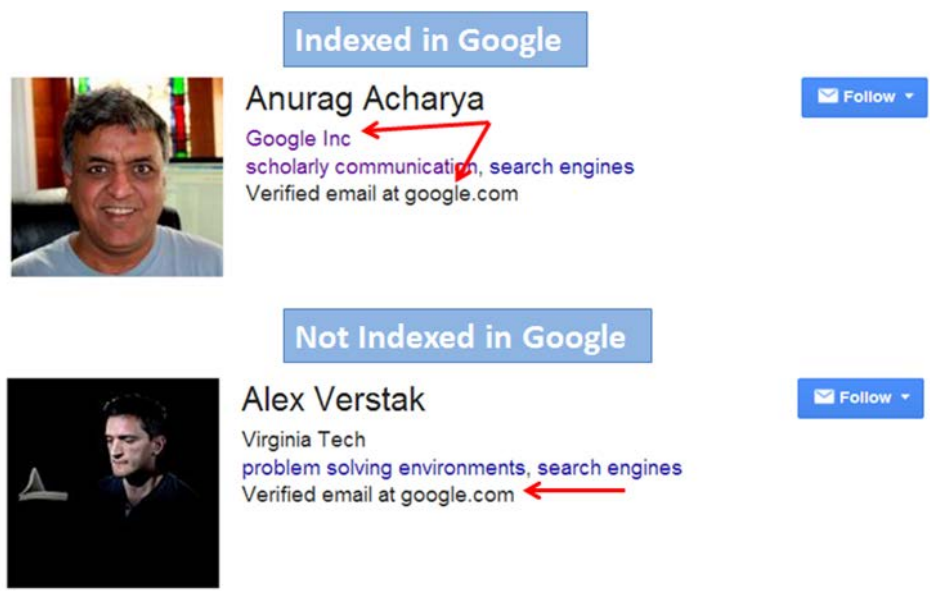

**Figure 10. Identification of normalized institutions at Google Inc.**

Let's analyze the Google case in depth, as a normalized institution within Google Scholar Citations. We identified a total of 1,043 scholars correctly assigned to the normalized entry "Google Inc". However, we found 1,116 profiles with a google.com verified email but not declaring any professional affiliation with the company (Figure 11, case A), and 465 profiles in which Google is declared as institutional affiliation, but they do not indicate an institutional email from the company (Figure 11, case B).

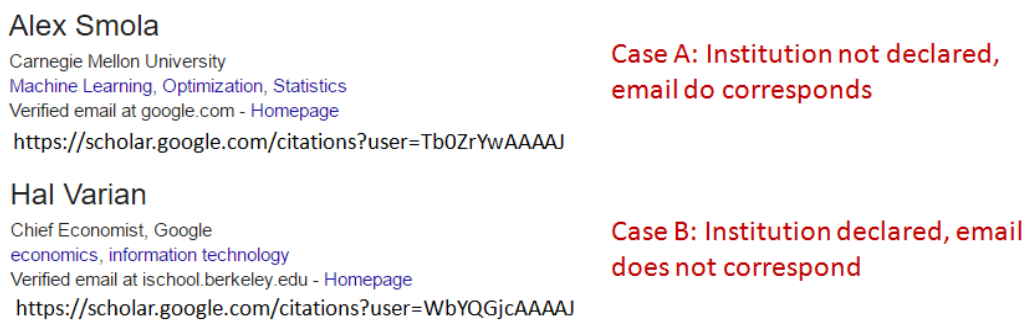

**Figure 11. Scholars not affiliated to Google Inc: affiliation/email inconsistency.**

If we assume that all scholars who have mentioned either an institutional affiliation to Google or have a google.com verified email are probably linked (currently or in the past) to this institution, approximately 538 scholars would have been excluded. The number of profiles in the normalized list would increase by more than 50% if these 538 were included.

**b) The case of Spanish universities**

We performed the same kind of searches with the Spanish universities, finding similar results. The number of profiles grouped under the normalized institutional affiliation for each university (institutional link) may be observed in Table 1. Additionally we supply the number of profiles recovered under a query by email web domain (for example, ugr.es), as well as the number of profiles under both web domain and institutional affiliation keyword queries (for example, "Universidad de Granada").

**Table I. Author profiles affiliated to Spanish universities in Google Scholar Citations**

| UNIVERSITY | IL | E | AF | EAWL | % |
|---|---|---|---|---|---|
| Univ. de Granada | **1,136** | 1,438 | 85 | 387 | 25.4 |
| Univ. Complutense de Madrid | **860** | 1,120 | 75 | 335 | 28.0 |
| Univ. Politécnica de Cataluña | **726** | 887 | 65 | 226 | 23.7 |
| Univ. Pompeu Fabra | **722** | 796 | 72 | 146 | 16.8 |
| Univ. de Barcelona | **674** | 834 | 208 | 368 | 35.3 |
| Univ. Autónoma de Barcelona | **636** | 787 | 99 | 250 | 28.2 |
| Univ. Politécnica de Valencia | **628** | 763 | 21 | 156 | 19.9 |
| Univ. de Sevilla | **613** | 825 | 66 | 278 | 31.2 |
| Univ. de Valencia | **592** | 780 | 66 | 254 | 30.0 |
| Univ. Politécnica de Madrid | **561** | 690 | 41 | 170 | 23.3 |
| Univ. Autónoma de Madrid | **522** | 656 | 49 | 183 | 26.0 |
| Univ. del País Vasco | **492** | 602 | 54 | 164 | 25.0 |
| Univ. Carlos III de Madrid | **469** | 549 | 49 | 129 | 21.6 |
| Univ. de Zaragoza | **453** | 583 | 38 | 168 | 27.1 |
| Univ. de Málaga | **438** | 543 | 30 | 135 | 23.6 |
| Univ. de Santiago de Compostela | **394** | 493 | 40 | 139 | 26.1 |
| Univ. de Alicante | **356** | 435 | 41 | 120 | 25.2 |
| Univ. de Murcia | **354** | 460 | 28 | 134 | 27.5 |
| Univ. Nacional de Educación a Distancia | **348** | 418 | 79 | 149 | 30.0 |
| Univ. de Cádiz | **321** | 402 | 18 | 99 | 23.6 |
| Univ. de Salamanca | **315** | 407 | 30 | 122 | 27.9 |
| Univ. de Castilla-La Mancha | **305** | 373 | 23 | 91 | 23.0 |
| Univ. de Oviedo | **302** | 365 | 39 | 102 | 25.2 |
| Univ. de Las Palmas de Gran Canaria | **286** | 349 | 16 | 79 | 21.6 |
| Univ. Jaime I | **281** | 330 | 9 | 58 | 17.1 |
| Univ. de Valladolid | **276** | 352 | 23 | 99 | 26.4 |
| Univ. de La Coruña | **273** | 327 | 5 | 59 | 17.8 |
| Univ. de Vigo | **258** | 317 | 21 | 80 | 23.7 |
| Univ. Rovira i Virgili | **254** | 290 | 38 | 74 | 22.6 |
| Univ. de La Laguna | **253** | 181 | 163 | 91 | 26.5 |
| Univ. de Extremadura | **244** | 306 | 13 | 75 | 23.5 |
| Univ. Rey Juan Carlos | **222** | 284 | 31 | 93 | 29.5 |
| Univ. de Gerona | **217** | 253 | 37 | 73 | 25.2 |
| Univ. de las Islas Baleares | **208** | 231 | 33 | 56 | 21.2 |
| Univ. Miguel Hernández de Elche | **198** | 252 | 14 | 68 | 25.6 |
| Univ. de Navarra | **195** | 237 | 49 | 91 | 31.8 |
| Univ. de Alcalá | **190** | 240 | 25 | 75 | 28.3 |
| Univ. de Córdoba | **179** | 232 | 21 | 74 | 29.2 |
| Univ. de Jaén | **178** | 214 | 8 | 44 | 19.8 |
| Univ. de Cantabria | **173** | 220 | 9 | 56 | 24.5 |
| Univ. Pablo de Olavide | **169** | 217 | 14 | 62 | 26.8 |
| Univ. de León | **135** | 179 | 4 | 48 | 26.2 |
| Univ. Oberta de Catalunya | **133** | 168 | 22 | 57 | 30.0 |

| | | | | |
|---|---|---|---|---|
| Univ. de Almería | **121** | 151 | 7 | 37 | 23.4 |
| Univ. de Lérida | **117** | 153 | 28 | 64 | 35.4 |
| Univ. de Huelva | **116** | 156 | 17 | 57 | 32.9 |
| Univ. de Deusto | **102** | 123 | 4 | 25 | 19.7 |
| Univ. Politécnica de Cartagena | **90** | 110 | 6 | 26 | 22.4 |
| Univ. Pública de Navarra | **90** | 108 | 1 | 19 | 17.4 |
| Univ. CEU Cardenal Herrera | **81** | 117 | 11 | 47 | 36.7 |
| Univ. CEU San Pablo | **81** | 117 | 7 | 43 | 34.7 |
| Univ. de Burgos | **62** | 80 | 7 | 25 | 28.7 |
| Univ. Ramon Llull | **59** | 66 | 0 | 7 | 10.6 |
| Univ. Católica San Antonio | **56** | 91 | 8 | 43 | 43.4 |
| Univ. Europea de Madrid | **54** | 61 | 7 | 14 | 20.6 |
| Univ. Internacional de La Rioja | **51** | 63 | 10 | 22 | 30.1 |
| Univ. de La Rioja | **45** | 51 | 5 | 11 | 19.6 |
| Univ. de Vich | **35** | 44 | 4 | 13 | 27.1 |
| Univ. Pontificia Comillas | **30** | 75 | 10 | 55 | 64.7 |
| IE University | **30** | 41 | 1 | 12 | 28.6 |
| Univ. Loyola Andalucía | **30** | 37 | 12 | 19 | 38.8 |
| Univ. Católica de Valencia San Vicente Mártir | **27** | 38 | 1 | 12 | 30.8 |
| Univ. Camilo José Cela | **24** | 30 | 8 | 14 | 36.8 |
| Univ. a Distancia de Madrid | **22** | 26 | 0 | 4 | 15.4 |
| Univ. Internacional de Cataluña | **19** | 31 | 5 | 17 | 47.2 |
| Univ. San Jorge | **17** | 20 | 1 | 4 | 19.0 |
| Univ. de Mondragón | **15** | 25 | 1 | 11 | 42.3 |
| Univ. Pontificia de Salamanca | **13** | 17 | 7 | 11 | 45.8 |
| Univ. Francisco de Vitoria | **12** | 15 | 4 | 7 | 36.8 |
| Univ. Internacional Isabel I de Castilla | **0** | 5 | 0 | 5 | 100 |
| Univ. Antonio de Nebrija | **0** | 21 | 4 | 25 | 100 |
| Univ. Abad Oliva CEU | **0** | 6 | 2 | 8 | 100 |
| Univ. Europea Miguel de Cervantes | **0** | 6 | 1 | 7 | 100 |
| Univ. Católica Santa Teresa de Jesús de Ávila | **0** | 3 | 0 | 3 | 100 |
| Univ. Internacional Valenciana | **0** | 3 | 0 | 3 | 100 |
| Univ. Eclesiástica San Dámaso | **0** | 2 | 2 | 4 | 100 |
| Univ. Europea de Canarias | **0** | 2 | 1 | 3 | 100 |
| Univ. Europea de Valencia | **0** | 2 | 1 | 3 | 100 |
| Univ. Europea del Atlántico | **0** | 2 | 0 | 2 | 100 |
| Univ. Internacional de Andalucía | **0** | 1 | 4 | 5 | 100 |
| Univ. Alfonso X el Sabio | **0** | 1 | 3 | 4 | 100 |
| Univ. Internacional Menéndez Pelayo | **0** | 0 | 0 | 0 | 0 |
| **TOTAL** | **17,938** | **22,285** | **2,061** | **6,408** | |

Google Scholar Citations groups 73.7% of all the scholars that could potentially have been linked to the institution by having an email or affiliation of the institution, and 80.5% of those with official email from the institution. There are therefore a very significant number of researchers without proper institutional linkage.

Additionally, we can observe up to 14 institutions (all of them private universities) with 0 authors included and no normalized affiliation (in some cases no verified emails were provided, or the name of the institution is not correct). We should mention the unexpected percentage obtained for the University of Barcelona: only 64.7% of authors included of the 1,042 potential authors who either entered the UB as their institution, or have a UB verified email.

To calibrate the extension of this problem we would have to conduct a comprehensive study. However, as a way to illustrate this problem, we have analyzed the sample of 82 existing Spanish universities (September 2015), obtaining that 13 of them (15.85%) didn't have a standardized GSC link.

## 6. Discussion and conclusions

The results obtained in this study allow us to respond negatively to all three research questions.

As regards RQ1, Google Scholar Citations is not able to generate an affiliation link to all institutions inserted in public Google Scholar profiles, suggesting a lack of automated identification procedures. Regarding RQ2, we find several errors that prove the current inability of the feature to generate a unique standardized name for all the institutions. Lastly, regarding RQ3, the analyses applied both to Google and to the whole Spanish academic system prove that the feature is not capable of grouping together all the researchers who work in a same institution

These findings do not compromise the usefulness of this new tool in scientific resource discovering processes, but jeopardize its direct use as a source for quantitative analyses unless data cleansing processes are applied beforehand. Unfortunately, although normalization tasks are less tedious than those required for data from Google Scholar, these are still needed.

This effect is not exclusive to Google Scholar Citations, however, and other platforms suffer similar problems. For example, Microsoft Academic exhibits more than 100 million publications not affiliated to any institution (Herrmannova & Knoth, 2016), an issue that may affect their institutional profiles to some extent (the purpose and features of these profiles are absolutely different from GSC profiles, however). ResearchGate avoids this problem by displaying a drop-down menu to select user affiliations. This solution makes ResearchGate profiles useful for quantitative analyses at the institutional level (Thelwall & Kousha, 2015). Despite this, the creation of internal units within universities (departments or Schools) is not subject to any control, thus opening the door to duplicates at those levels.

We should not be surprised about inconsistencies and errors made by Google Scholar Citations, because the task of authority control is devilishly complex. These errors are the natural consequence of a product conceived from the "laissez faire laissez passer" principle (Delgado López-Cózar, 2014): "Total freedom is left to authors in order to set out the profile to their own taste and benefit, from their personal, professional and subject identification to the possibility of linking bibliographic production they deem appropriate or even selecting the mode of profile updates (automatic or manually). The author can add, edit and delete bibliographic records as he/she pleases. Google Scholar doesn't in any way verify the truthfulness and accuracy of the data and leaves to the author the full responsibility for what is shown."

This complete autonomy granted to the author reflects Google's philosophy: a collection of user-oriented information; created by and for the user. A clever way not only to reduce the costs derived from information processing and management, but also to make the user "work for the system". Nonetheless, avoiding authority control is an approach – as we can see – that has its own risks.

This study attempts to analyze the cleaned-up and structured data available in Google Scholar Citations, and more specifically, the data related to institutional affiliations, aiming to the effectiveness of the new affiliation link tool for grouping all authors with a public profile belonging to one institution. Therefore, other bibliographic and citation errors (either in Google Scholar or in Google Scholar Citations) are not considered in this study. Google Scholar general search engine does not capture information about authors' institutional affiliation. This data is provided directly by users when editing their academic profiles. For this reason, the affiliation tool effectiveness does not depend on the Google Scholar operating.

In this sense, all errors identified and classified in this work depend on the one hand on the personal information provided by users, and on the other hand on how the system performs when the information is incorrect (e.g., misspelled, invented, etc.), insufficient (e.g., non-standardized acronyms, names for internal units, etc.) or even when it is correct but showing special particularities (e.g., double affiliations, mixed affiliations, unknown affiliations, etc.). All cases shown in the results section exemplify the current functioning of the affiliation tool, with its advantages and disadvantages.

While a large number of the errors identified in this work are common to other general institution name disambiguation problems in other bibliographic databases (spelling, translation errors, disambiguation problems, wrong normalized name, incorrect linking), some others are particular to the Google Scholar Citation's affiliation tool (multiple official academic web domains, lack of precision of the search box, complex institutions, multiple institutions, internal affiliations). In any case, this work intended solely to highlight the institutional name mistakes that this Google Scholar tool actually exhibit, with the general purpose of preventing users from the information offered by the tool.

Google Scholar Citations' affiliation tool was launched in August 2015, and to the best of our knowledge, no other empirical study has treated this issue to date. Therefore, there is no way to compare the original results obtained with any other similar work. Some of the general pros (cleaned data) and cons (lack of normalization) about Google Scholar Citations profiles were introduced by Jacsó (2012b) and Delgado López-Cózar (2014), long before the affiliation tool was conceived.

The main findings of this study show that currently, the affiliation link provided by Google Scholar Citation profiles is unable to group all existing authors with a public profile created belonging to one institution. The consequences of this are two-fold. First, bibliometric studies should not use information from GSC' institutional lists directly (without prior processing) to perform quantitative analyses, whether about presence of institutions (members with created profiles), or their impact (global citations received by one institution). Secondly, professionals (especially those related to university policies) should not use these institutional lists for marketing or self-promotion purposes, or to develop university rankings. This could lead non-experts to believe these institutional rankings are exhaustive, when they truly are not.

We are aware that this study only focuses on a limited number of queries, and that is centered on one specific country. New cases may arise if different national academic systems are studied. However, the identified errors have been empirically proven to exist.

Lastly we deem it necessary to formulate a series of recommendations to profile users and other actors involved with this product, which are only aimed at improving and to make the most of it.

*To Google Scholar*

- Improve the personal information form users have to fill when creating a profile, providing fields with information about the institution and the country. The system might suggest the normalized name of the institution, and the correct acronym.
- Allow authors to include several institutions, since it is very common for researchers to work in various different institutions throughout their academic life. For example, an open box for past institutions and years of affiliation.

In order to help users standardize institutions, Google can use the technique it already uses in the general search engine, in which as the user is typing the name of an institution, the system automatically offers different standardized options to select the desired one. In the case of countries, a dropdown list could be used.

To avoid omissions (deliberate or not), the use of mandatory fields may be adopted. In this way, many problems are fixed from the beginning, which is the best option to develop an accurate and comprehensive authorities' control base in an open system like this. Google has never focused (neither the general nor academic search engine) in such controlled-vocabulary, but has rather favored natural language searching. However, the combination of these two techniques may help feeding the system with clean data and improving its precision.

*To authors*

We warn authors about the potential disadvantages of not including the name of the institution they are affiliated to and verifying their institutional email, a decision that will automatically exclude them from the institutional lists. Moreover, they must not only indicate the name of the institution, but they should do it using the normalized nomenclature.

The attention paid to this subject reflects that institutional names should not be translated, but instead expressed in the original language (Universidad de Jaén instead of University of Jaen).

Although nowadays translating names to English is a widespread practice to facilitate international visibility, and to prevent the proliferation of variants, this practice is recommended only when the center has an accepted and standardized name in this language.

*To institutions*

We advise institutions not to develop research evaluation products directly from Google Scholar Citations' institutional lists without first conducting a thorough identification of all researchers belonging to their institution, linked through the Google Scholar's institutional affiliation link or not. Institutions should develop specific policies to encourage its staff to use its corporate name in a standardized way.

We do hope the profound changes experienced by Google (now Alphabet) in summer 2015 do not pose any threat to Google Scholar so that it can keep growing and improving, not only as an academic search engine but also as a valuable source of data for bibliometric studies.